\documentclass[aps, prl, reprint, superscriptaddress]{revtex4-1}
\pdfoutput=1 
\usepackage{graphicx}
\usepackage{color,amsmath}

\usepackage[normalem]{ulem}

\usepackage[colorlinks,linkcolor=blue,citecolor=blue,urlcolor=blue]{hyperref}

\usepackage{comment}
\begin{document}

\title{Non-local polarization feedback in a fractional quantum Hall ferromagnet}

\author{Szymon Hennel}
\email{hennels@phys.ethz.ch}
\affiliation{Solid State Physics Laboratory, ETH Z\"{u}rich, 8093 Z\"{u}rich, Switzerland}

\author{Beat A. Braem}
\affiliation{Solid State Physics Laboratory, ETH Z\"{u}rich, 8093 Z\"{u}rich, Switzerland}

\author{Stephan Baer}
\affiliation{Solid State Physics Laboratory, ETH Z\"{u}rich, 8093 Z\"{u}rich, Switzerland}

\author{Lars Tiemann}
\affiliation{Solid State Physics Laboratory, ETH Z\"{u}rich, 8093 Z\"{u}rich, Switzerland}

\author{Pirouz Sohi}
\affiliation{Solid State Physics Laboratory, ETH Z\"{u}rich, 8093 Z\"{u}rich, Switzerland}

\author{Dominik Wehrli}
\affiliation{Solid State Physics Laboratory, ETH Z\"{u}rich, 8093 Z\"{u}rich, Switzerland}

\author{Andrea Hofmann}
\affiliation{Solid State Physics Laboratory, ETH Z\"{u}rich, 8093 Z\"{u}rich, Switzerland}

\author{Christian Reichl}
\affiliation{Solid State Physics Laboratory, ETH Z\"{u}rich, 8093 Z\"{u}rich, Switzerland}

\author{Werner Wegscheider}
\affiliation{Solid State Physics Laboratory, ETH Z\"{u}rich, 8093 Z\"{u}rich, Switzerland}

\author{Clemens R\"ossler}
\affiliation{Solid State Physics Laboratory, ETH Z\"{u}rich, 8093 Z\"{u}rich, Switzerland}

\author{Thomas Ihn}
\affiliation{Solid State Physics Laboratory, ETH Z\"{u}rich, 8093 Z\"{u}rich, Switzerland}

\author{Klaus Ensslin}
\affiliation{Solid State Physics Laboratory, ETH Z\"{u}rich, 8093 Z\"{u}rich, Switzerland}

\author{Mark S. Rudner}
\affiliation{Niels Bohr International Academy and Center for Quantum Devices, Niels Bohr Institute, University of Copenhagen, 2100 Copenhagen, Denmark}

\author{Bernd Rosenow}
\affiliation{Institut f\"ur Theoretische Physik, Universit\"at Leipzig, D-04009 Leipzig, Germany}

\date{\today}

\begin{abstract}
In a quantum Hall ferromagnet, the spin polarization of the two-dimensional electron system can be dynamically transferred to nuclear spins in its vicinity through the hyperfine interaction. The resulting nuclear field typically acts back locally, modifying the local electronic Zeeman energy. Here we report a non-local effect arising from the interplay between nuclear polarization and the spatial structure of electronic domains in a $\nu=2/3$ fractional quantum Hall state. In our experiments, we use a quantum point contact to locally control and probe the domain structure of different spin configurations emerging at the spin phase transition. Feedback between nuclear and electronic degrees of freedom gives rise to memristive behavior, where electronic transport through the quantum point contact depends on the history of current flow. We propose a model for this effect which suggests a novel route to studying edge states in fractional quantum Hall systems and may account for so-far unexplained oscillatory electronic-transport features observed in previous studies.
\end{abstract}

%\keywords{}

\maketitle

In a wide variety of low-dimensional electronic systems, the electron spin degrees of freedom are coupled to a bath of nuclear spins residing in the atomic cores of the host lattice. If the electronic system is driven out of equilibrium, this coupling may cause the nuclear spins to become {\it dynamically polarized,} in turn feeding back to locally affect the electronic state. In the simplest case, the result is decoherence or depolarization of the electronic spin state. More interestingly, complex dynamical phenomena arising from local feedback effects have also been reported~\cite{Latta2009,Vink2009, ono2004}. Here we report an electronic--nuclear feedback mechanism that is {\it non-local} in nature: the spatial regions where polarization is created and those where the polarization rates are controlled are spatially separated, with feedback between these processes mediated through nuclear spin diffusion.

We employed the particular properties of the $\nu=2/3$ fractional quantum Hall (FQH) state~\cite{eisenstein_evidence_1990}, which undergoes a spin phase transition when two composite fermion~\cite{jain_composite-fermion_1989} levels with opposite spin quantum numbers cross at the chemical potential. Near the phase transition point, the electron system breaks up into micrometer-sized domains ~\cite{hayakawa_real-space_2013}, with the spin--orbital configuration of the FQH state taking the role of a pseudospin quantum number ~\cite{smet_ising_2001, eom_quantum_2000}, and electronic transport is accompanied by dynamic polarization of nuclear spins ~\cite{paget_low_1977, wald_local_1994, kronmuller_new_1998, kronmuller_new_1999, kraus_quantum_2002, kou_dynamic_2010}. As a sensitive probe of nuclear magnetization, we use a quantum point contact (QPC) with local filling factor $\nu=2/3$ ~\cite{yusa_controlled_2005}. At constant magnetic field, reduced carrier density favors the polarized ground state ~\cite{hayakawa_real-space_2013}, suggesting the use of a QPC for the control of domain morphology at the level of single domains. 

\begin{figure}
\includegraphics[width=\columnwidth]{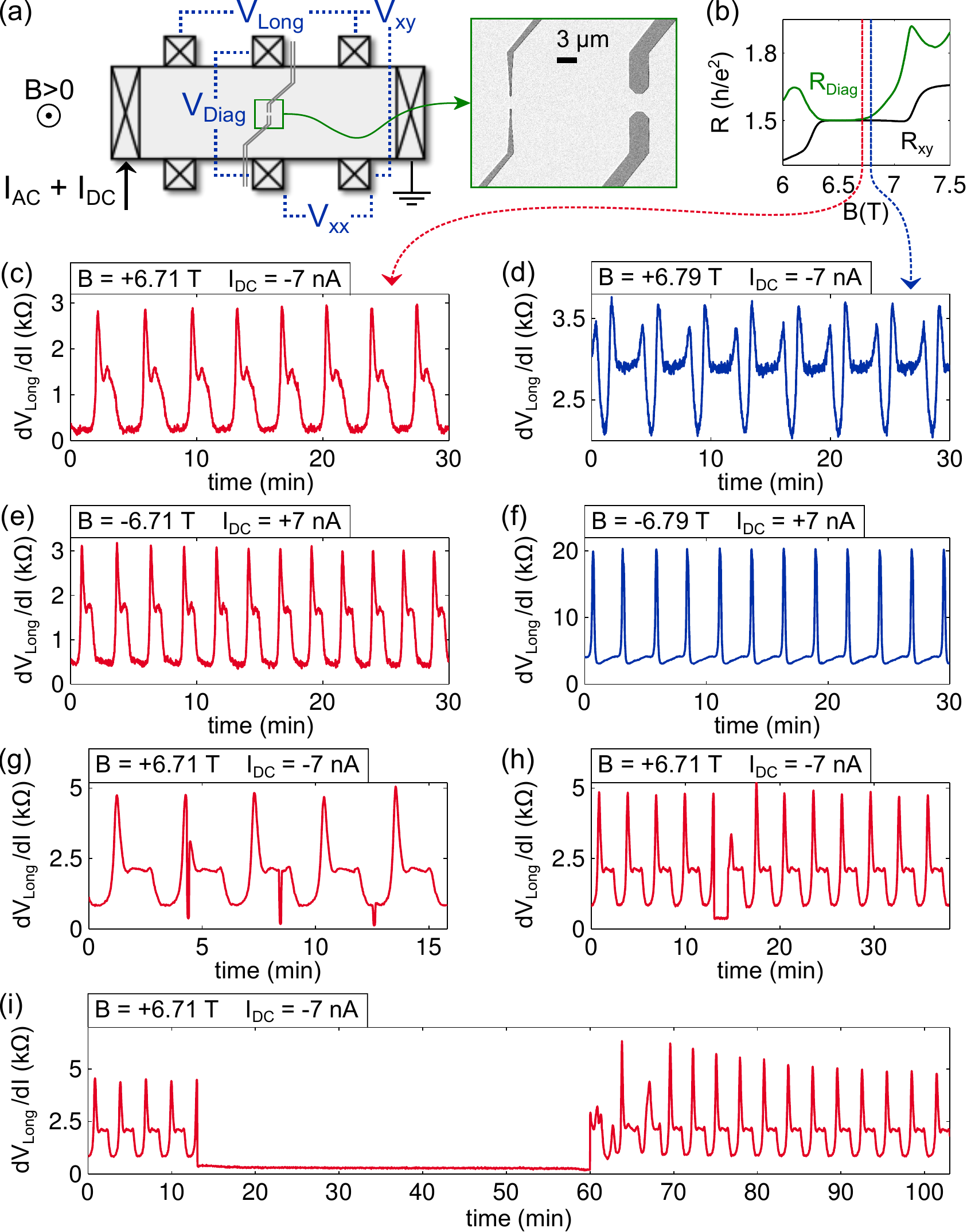}
\caption{\textbf{(a)} Schematic of the sample. Crosses represent Ohmic contacts, voltage probes are indicated by dotted lines. Top-gate-defined QPCs with gaps ranging from $800$ nm to $2.8$ $\mu$m in the central region, two are shown in the inset (SEM image). \textbf{(b)} Bulk Hall resistance and diagonal resistance measured across a QPC with 1800 nm lithographic gap at a bath temperature of 70 mK with $-0.75$ V applied to the corresponding gates. Dashed lines indicate $B=6.71$ T and $B=6.79$ T. \textbf{(c) to (f)} Time traces of the longitudinal differential resistance. The bias current and the magnetic field are indicated on top of each graph. \textbf{(g)} The DC bias current is set to zero for 10 seconds, at three different points in consecutive periods. The differential longitudinal resistance drops, and oscillations continue after the original bias is restored. \textbf{(h)} The DC bias is interrupted for 100 seconds. \textbf{(i)} Interruption of the DC bias for 57 minutes, a time span of aperiodic fluctuations is subsequently observed.}
\label{fig1}
\end{figure}

We observe pronounced time-dependent oscillations of the resistance measured across the QPC when a DC bias current is applied to the sample at magnetic fields corresponding to a flank of the $\nu=2/3$ plateau in the effective QPC filling factor ($R_\text{Diag}$ in Fig.~\ref{fig1}(b)). We obtain similar results in several different QPCs, establishing the self-sustained resistance oscillations reported by Yusa {\it et al.}~\cite{yusa_self-sustaining_2004} as a generic phenomenon. We employ timed nuclear magnetic resonance (NMR) pulses to provide a new window into the relation between sharp features in current and spin dynamics and propose a model involving the interplay of dynamic nuclear polarization (DNP) with the spatial structure of electronic spin-polarized and unpolarized $\nu = 2/3$ domains.

Our sample is described in Fig.~\ref{fig1}(a). Several QPCs were fabricated by Ti/Au evaporation on a photolithographically defined $400\ \mu$m wide Hall bar in a single-side doped GaAs/AlGaAs heterostructure. For each measurement, a single QPC was defined by application of a negative voltage to the corresponding gates. The 2DEG with density $n_\text{s} \approx 1.0\times 10^{11}$ cm$^{-2}$ and mobility $\mu \approx 4 \times 10^{6}$ cm$^2$V$^{-1}$s$^{-1}$ at $4.2$ K resided $310$ nm beneath the sample's surface. Self-consistent calculations yield a wave-function extent of $30$ nm in growth direction. In the vicinity of $\nu=2/3$ we observe the distinct signatures of a spin phase transition~\cite{supplement}. Four-terminal measurements of differential resistance were performed in a dilution refrigerator using standard lock-in techniques with an excitation of $100$ pA at 24.44 Hz. AC and DC currents were supplied through 1 G$\Omega$ resistors. We studied two samples processed from the same wafer in three cool-down cycles. For time-dependent measurements, one data point per second was acquired.

A typical time trace of the differential longitudinal resistance in the steady state reached after application of a DC current is shown in Fig.~\ref{fig1}(c). The oscillations remain stable in measurements spanning up to 30 hours. Data taken at intervals of several weeks show oscillations of identical shape, periodicity and amplitude for identical magnetic-field value, current bias and temperature. The phenomenon is sensitive to small differences in magnetic field (compare Fig.~\ref{fig1}(c) and Fig.~\ref{fig1}(d)). Oscillations are not symmetric under inversion of both magnetic field and bias direction, with more pronounced asymmetry away from the middle of the plateau in diagonal resistance (see Fig.~\ref{fig1} (c) to (f)). In contrast to Ref. \cite{yusa_self-sustaining_2004}, we do not observe any qualitative difference between sources of constant current and sources of constant voltage. Oscillations are also observed in DC-only measurements, and are damped by AC current levels exceeding 500 pA independently with no dependence on lock-in frequency. 

The dynamics of our sample depend on the history of current flow. Once the DC bias current is set to zero, the differential resistance drops abruptly. On restoration of the initial DC bias, the oscillations continue from the point where current flow was suspended, provided that the interval is shorter than or on the order of the oscillation period (which is here several minutes). Memory of past current flow deteriorates for time intervals of zero-DC bias that are substantially longer than one oscillation period. In those cases, random fluctuations are observed before the oscillatory steady-state is re-established. Memristive effects in our devices are summarized in Fig.~\ref{fig1}(g) to (i). 
%%%%%%%%%%%%%%%%%%%%%%%%%%%%%%%%%%%%%%%%%%%%%%%%%%%%%%%%%%%%%%%%%%
%%%%%%%%%%%%%%%%%%%%%%%%%%%%%%%%%%%%%%%%%%%%%%%%%%%%%%%%%%%%%%%%%%

\begin{figure}
\includegraphics[width=\columnwidth]{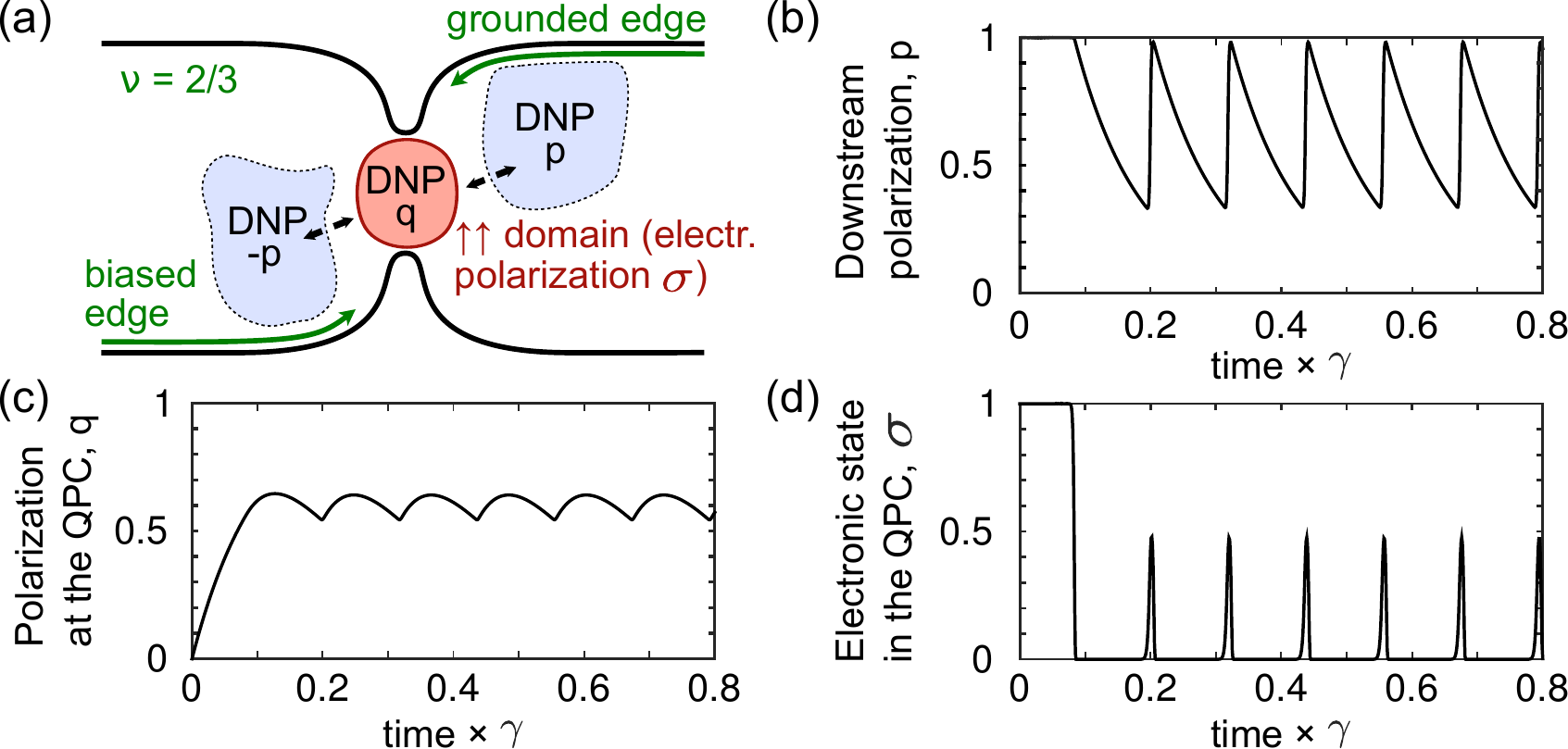}
\caption{Model for self-oscillations driven by DNP.  
\textbf{(a)} A spin-polarized electron domain resides in the constriction, with degree of polarization $\sigma$, resulting in different transmissions for quasiparticles with up and down spins. Spin filtering leads to spin accumulation outside the constriction region, which is transferred to nuclei with polarization $p$ and $-p$. Through nuclear spin diffusion, DNP ($q$) builds up in the constriction, destabilizing the polarized domain.
When the domain shrinks the spin filtering effect stops; after the DNP diffuses away the domain reforms and oscillations result.
\textbf{(b) to (d)} Time-dependent oscillations produced by the model system (\ref{eq:Model}).
Parameter values are given in the text.
}
\label{fig2}
\end{figure}

We propose that the observed behavior results of the interplay between DNP and the spatial structure of electronic spin-polarized and unpolarized $\nu = 2/3$ domains (see Fig.~\ref{fig2}(a)). For the range of fields where oscillations are observed, we expect that the bulk is primarily comprised of the unpolarized $\nu=2/3$ state. Noting that the spin-polarized state is favored at lower densities at constant magnetic field, we surmise that in equilibrium a spin-polarized domain is formed inside the constriction. For an unpolarized impinging current, the polarized domain inside the constriction acts as a spin filter. For an upstream direction of spin-mode propagation, as predicted e.g. in \cite{MooreHaldane}, spin accumulation will occur in the regions shown in Fig.~\ref{fig2}(a), and hyperfine exchange will mediate a polarization transfer to the nuclear-spin subsystem. As a result, two spatial regions of opposite nuclear polarization (denoted $p$ and $-p$) emerge in the vicinity of the QPC. A part of the resulting DNP spreads to the constriction through isotropic nuclear spin diffusion, a slow process accounting naturally for the long timescales observed ($\sim$100 seconds are expected for diffusion across $\sim$1 $\mu$m). If the net nuclear polarization in the constriction ($q$) is positive, the corresponding Overhauser field {\it destabilizes} the polarized electronic domain, suppressing the spin-filtering effect. Once the spin-diffusion-mediated nuclear polarization at the constriction fades away, the polarized $\nu = 2/3$ domain may form again, initiating a new oscillation cycle.
%%%%%%%%%%%%%%%%%%%%%%%%%%%%%%%%%%%%%%%%%%%%%%%%%%%%%%%%%%%%%%%%%%
%%%%%%%%%%%%%%%%%%%%%%%%%%%%%%%%%%%%%%%%%%%%%%%%%%%%%%%%%%%%%%%%%%

\begin{figure}
\includegraphics[width=\columnwidth]{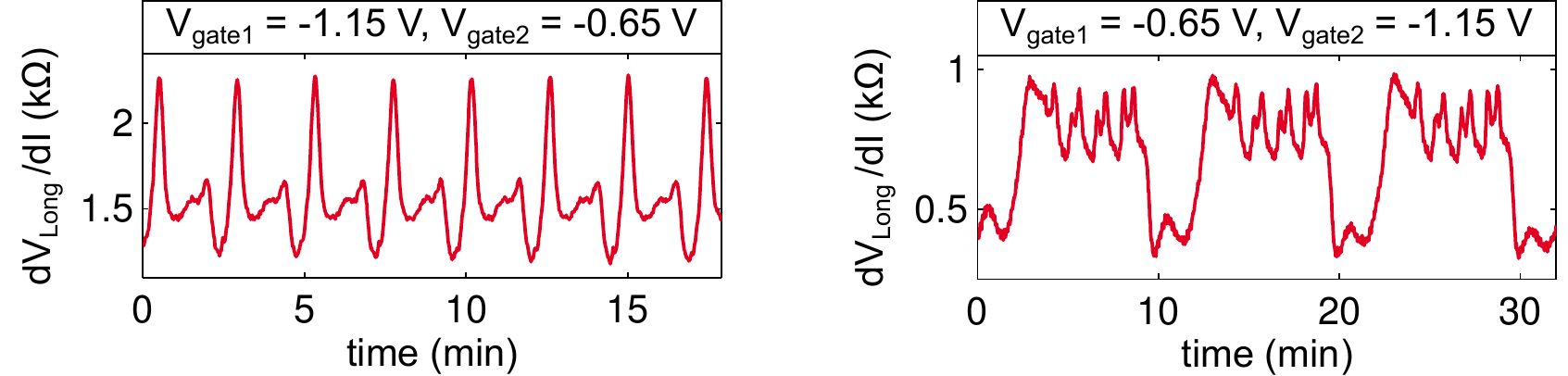}
\caption{Influence of asymmetric gate voltage configurations (indicated on top) on the differential longitudinal resistance measured across a QPC with 2000 nm gap at B=$6.1$ T with a bias current of $-10$ nA.}
\label{fig3}
\end{figure}

The oscillations are sensitive to a distortion and shift of the QPC confinement potential resulting from an asymmetric gate voltage configuration, confirming that the underlying dynamics unfold in close vicinity of the constriction. The effect of displacing the QPC by $\sim100$ nm while keeping its zero-field transmission constant is shown in Figures \ref{fig3}(a) and (b). 
%%%%%%%%%%%%%%%%%%%%%%%%%%%%%%%%%%%%%%%%%%%%%%%%%%%%%%%%%%%%%%%%%%
%%%%%%%%%%%%%%%%%%%%%%%%%%%%%%%%%%%%%%%%%%%%%%%%%%%%%%%%%%%%%%%%%%

\begin{figure}
\includegraphics[width=\columnwidth]{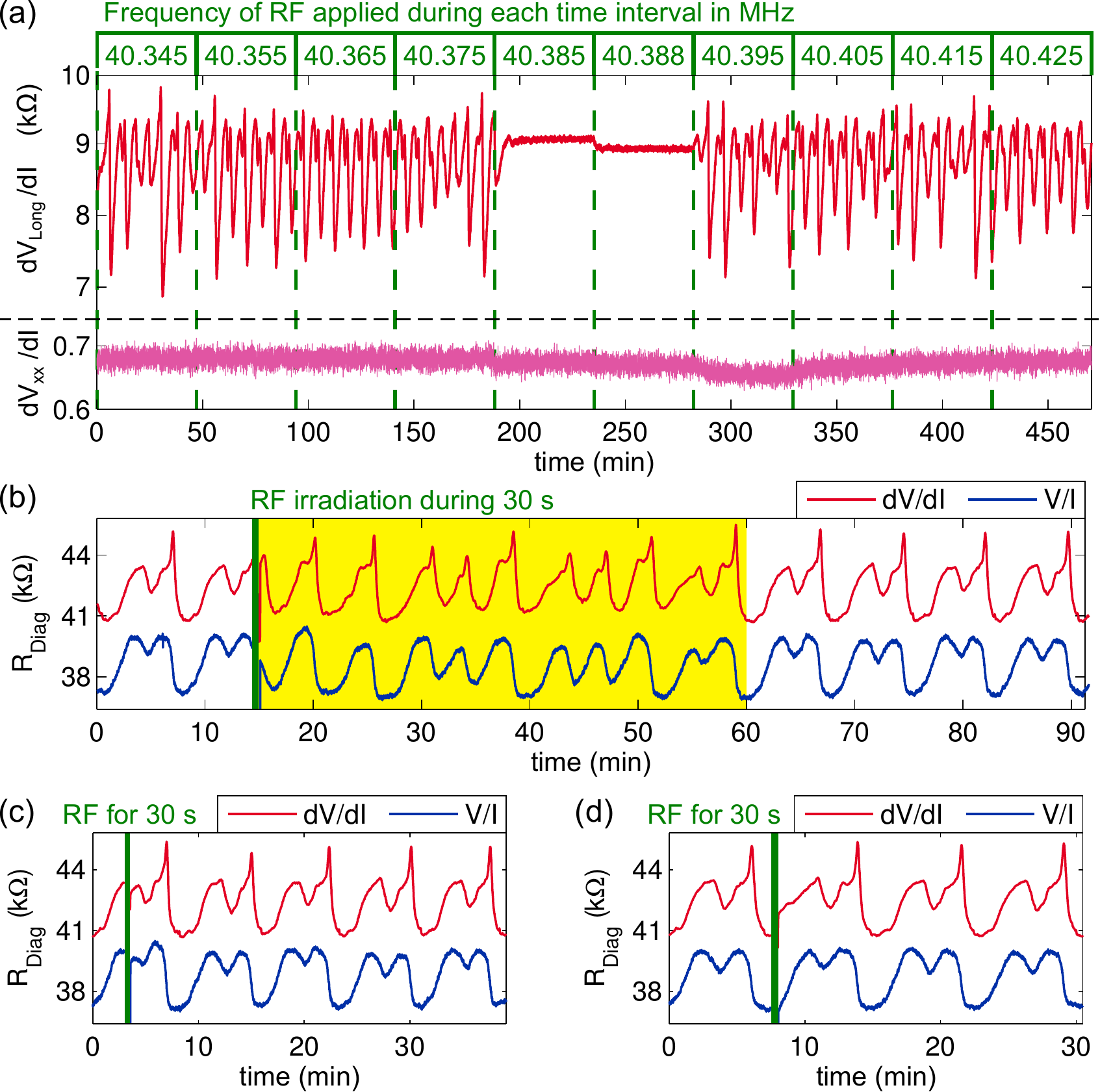}
\caption{\textbf{(a)} \textbf{Top} (red curve): differential resistance across a QPC with $2400$ nm gap measured while irradiating the sample with RF. The RF frequency is incremented at regular time intervals. Oscillations driven by a DC current of $9$ nA are suppressed when the RF frequency is close to the resonance frequency of $^{75}$As. \textbf{Bottom} (pink curve): simultaneously measured differential resistance in the bulk of the sample. Note the different scale on the y-axis. \textbf{(b)} Response to short irradiation at resonance frequency of $^{75}$As, $^{69}$Ga and $^{71}$Ga for 10 seconds each during the time interval indicated by a green bar, coinciding with a transient nonlinearity in the current-voltage characteristics. Aperiodic fluctuations of $R_\text{Diag}$ are observed during the time highlighted in yellow. \textbf{(c)} and \textbf{(d)} Response to the same RF sequence as in (b) applied when DC and differential resistances evolve similarly.}
\label{fig4}
\end{figure}

We confirmed the relevance of nuclear polarization through NMR experiments~\cite{kronmuller_new_1999}. We put a wire loop around our sample and applied an RF signal of $-20$ dBmW power, generating an AC field perpendicular to the external magnetic field. We extract nuclear relaxation times on the order of $80$ s from the evolution of $V_\text{Long}$ at zero DC bias after a continuous RF is set away from resonance at constant power. A continuous wave at the resonance frequency of $^{75}$As, $^{69}$Ga or $^{71}$Ga fully suppresses the oscillations, as shown in Fig.~\ref{fig4}(a) for $^{75}$As. The weak resonance observed in $\text{d}V_\text{xx}/\text{d}I$ at a slightly higher frequency is consistent with the presence of an unpolarized bulk 2DEG with a vanishing Knight shift in the bulk. Irradiation with resonant RF for short times affected the oscillations only if coinciding with a transient nonlinearity in the current--voltage characteristics, which we detected by comparing differential resistance to DC resistance (see Fig.~\ref{fig4}(b)). Irradiating the sample at a time coinciding with a sharp maximum in differential resistance resulted in aperiodic fluctuations for times exceeding 30 minutes, until the oscillatory steady state was restored. In contrast, no effect on the dynamics was observed when the RF burst is applied at a point where differential resistance and DC resistance follow the same time evolution (Fig \ref{fig4}(c) and (d)). 

We interpret our data considering the Knight shift in the QPC. The resonance frequency used for the RF bursts was detected in $R_\text{Long}$ in the absence of a DC bias. We observed a shift on the order of $5$ kHz with respect to the resonance detected in the bulk voltage $R_\text{xx}$ (the FWHM of the resonances is $10$ kHz). We thus expect the effect of RF bursts to be contingent on the electronic state of the QPC. We conclude that the sharp maxima in the differential resistance indeed indicate the nucleation of a short-lived polarized domain in the QPC. This transient nonlinearity suggests the presence of a domain morphology for a short time on the scale of the oscillations \cite{eom_quantum_2000}.
%%%%%%%%%%%%%%%%%%%%%%%%%%%%%%%%%%%%%%%%%%%%%%%%%%%%%%%%%%%%%%%%%%
%%%%%%%%%%%%%%%%%%%%%%%%%%%%%%%%%%%%%%%%%%%%%%%%%%%%%%%%%%%%%%%%%%

The feedback mechanism described in Fig. \ref{fig2}(a) is captured by the following dynamical system for the nuclear polarizations $p$ and $q$, and a variable $\sigma$ describing the electronic state in the constriction (ranging from $\sigma=0$ for an unpolarized state to $\sigma=1$  for a fully developed polarized domain): 
\begin{eqnarray}
\nonumber \dot{p} & = & f_0 (1 - p) \sigma^\alpha \ - \ \gamma_\perp p \ + \ \gamma q \label{pdot.eq}\\ %[.5cm]
\nonumber \dot{q} & = & \gamma p - \gamma_\perp q  \label{qdot.eq}  \\ %[.5cm]
\dot{\sigma} & = & (1 - \sigma) g( q_0 - q) -  \sigma g(q - q_0 - \Delta). \label{eq:Model}
\end{eqnarray}
Here $f_0$ is proportional to the bias current, $\alpha$ characterizes the spin-filtering efficiency of the polarized domain, $\gamma_\perp$ is a relaxation rate, $\gamma$ is a coupling constant capturing (slow) DNP diffusion between remote regions and the constriction, and $g$ is a step-like function which grows steeply when its argument crosses zero describing the growth and disappearance of the polarized domain on a fast time scale associated with electronic dynamics. Above the threshold $q_0$, the polarized domain is destabilized. The parameter $\Delta$ characterizes hysteresis of domain formation; once the domain disappears, the energy barrier to form a new domain requires the local DNP to decay to an amount $\Delta$ below $q_0$ ~\cite{hayakawa_real-space_2013, Verdene}. Such hysteresis appears to be crucial for obtaining stable self-oscillations.

The dynamical system (\ref{eq:Model}) yields stable self-oscillations over a wide range of physically reasonable parameter values. Representative results are shown in Fig.~\ref{fig2}(d). Here we used $g(q) = {1 \over 2} g_0 (1 - \sigma) \left[ 1 + \tanh \eta( q_0 - q) \right] - {1 \over 2} g_0 \sigma \left[1 + \tanh \eta (q - q_0 - \Delta)\right]$, with $\gamma_\perp/\gamma = 10$, $f_0/\gamma = 10^4$, $g_0/\gamma = 10^6$,  $q_0 = 0.05$,  $\Delta = 0.01$, $\eta = 10^3$, and $\alpha = 3$. These parameter values reflect the hierarchy of time scales discussed above. We generally find distinctly non-sinusoidal behavior, with the polarized state appearing only during sharp spike-like intervals (see Fig.~\ref{fig2}). The spikes in $\sigma(t)$ are reminiscent of the sharp features observed in differential resistance in the experiment (see, for example, Fig.~\ref{fig1}(f)). Furthermore, it seems natural to associate the degree of polarization with resistance, due to the same filtering effect that leads to DNP. The observation of multiple peaks within a single oscillation period, as shown in Fig.~\ref{fig1}(d), suggests a more complex domain morphology in the constriction. We have investigated possible simplifications of the model, eliminating the dynamical variables $\sigma$ and $\alpha$. If the source term $f_0 \sigma^\alpha$ is replaced by an arbitrary positive function $f(q)$, the dynamical system has only stable fixed points and no oscillations occur, even when a time delay in the coupling between $p$ and $q$ through spin diffusion is included.

\begin{figure}
\includegraphics[width=\columnwidth]{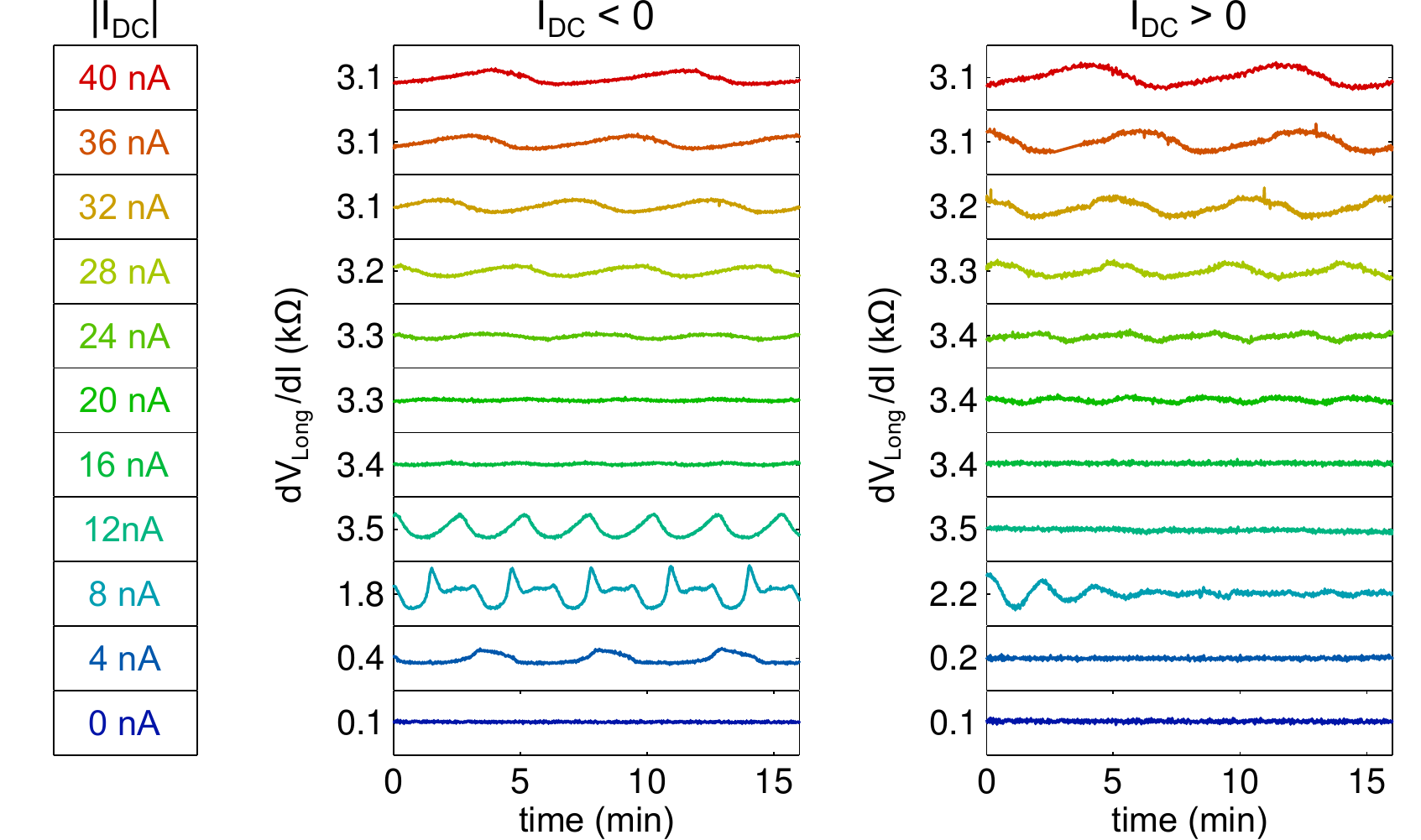}
\caption{Bias dependence of resistance oscillations in a QPC with 1800 nm gap at B=$6.745$~T. The absolute value of DC bias is indicated on the left. Each y-axis spans $3.5$~k$\Omega$ and a single tick indicates the average resistance.}
\label{fig5}
\end{figure}

We stress that an oscillatory solution is found only for the direction of DC bias which results in a positive polarization in the region with stronger diffusive coupling to the QPC. The bias dependence in a QPC with 1800 nm gap is shown in Fig.~\ref{fig5}. Sustained oscillations are observed when the DC current is on the order of $10$ nA, and only for one direction of the bias. Oscillations are suppressed as the bias is increased. A qualitatively different phenomenon is observed as oscillations re-emerge beyond $20$ nA without dependence on bias direction. At high bias, a shift of the QPC potential has no effect, suggesting that only outer regions of the QPC are involved. 
%%%%%%%%%%%%%%%%%%%%%%%%%%%%%%%%%%%%%%%%%%%%%%%%%%%%%%%%%%%%%%%%%%
%%%%%%%%%%%%%%%%%%%%%%%%%%%%%%%%%%%%%%%%%%%%%%%%%%%%%%%%%%%%%%%%%%

All data discussed above were acquired with both the constriction and the bulk 2DEG in the $\nu=2/3$ state. When the QPC was tuned to a lower effective filling factor by adjusting the gate voltage at constant magnetic field, only the symmetric high-bias regime was observed. In narrower QPCs, where the effective filling factor is $\nu=2/3$ at $B\simeq4$~T, we observed oscillations driven by a low bias, with no re-entrant oscillations as the bias was increased up to 200 nA. At such low fields the $\nu=2/3$ state in the QPC is expected to be unpolarized, whereas the bulk is in the fully polarized $\nu=1$ state. A different phenomenology might thus be at play. We did not observe oscillations in configurations involving only the  $\nu=1$ or $\nu=1/3$ states~\cite{supplement}.

In conclusion, we have presented a case where the effect of DNP goes beyond mere local depolarization. The memory of current flow through a domain of the polarized $\nu=2/3$ FQH state pinned by the confinement potential of a QPC is stored in nuclear spins, and diffusion of nuclear polarization in turn acts back on the electronic domain structure, resulting in persistent and precisely regular resistance oscillations. The model we propose relates the propagation direction of spin modes to spatial regions where spin is accumulated, suggesting novel possibilities for the study of edge reconstruction in FQH states that can undergo a spin-phase transition. More importantly perhaps, our work constitutes a first step towards a description of DNP at the microscopic scale and may have important implications for resistively detected NMR, a method which is poorly understood and yet widely used in the study of coupled electron-nuclear spin systems.
%%%%%%%%%%%%%%%%%%%%%%%%%%%%%%%%%%%%%%%%%%%%%%%%%%%%%%%%%%%%%%%%%%
\begin{acknowledgments}
The authors wish to thank J. Smet and S. H. Simon for useful discussions and gratefully acknowledge the support of the ETH FIRST laboratory and the financial support of the Swiss Science Foundation (Schweizerischer Nationalfonds, NCCR QSIT).
\end{acknowledgments}


\end{document}